\documentclass[preprint,aps,12pt,showpacs,nofootinbib,tightenlines,amsmath,amssymb]{revtex4}
\usepackage{amsmath}
\usepackage{graphicx}
\usepackage{amssymb}
\usepackage{color}

\newcommand{\p}{\partial}

\newcommand{\MeV}{\;\text{MeV}}
\newcommand{\GeV}{\;\text{GeV}}

\begin{document}
\def\intdk{\int\frac{d^4k}{(2\pi)^4}}
\def\sla{\hspace{-0.17cm}\slash}
\hfill

\title{Thermal Mass Spectra of Scalar and Pseudo-Scalar Mesons in IR-improved Soft-Wall AdS/QCD Model with Finite Chemical Potential}

\author{Ling-Xiao Cui}\email{clxyx@itp.ac.cn}

\author{Yue-Liang Wu}\email{ylwu@itp.ac.cn}

\affiliation{State Key Laboratory of Theoretical Physics(SKLTP)\\
Kavli Institute for Theoretical Physics China (KITPC) \\
Institute of Theoretical Physics, UCAS \\
Chinese Academy of Sciences, Beijing, 100190, China}

\date{\today}

\begin{abstract}
The thermal mass spectra of scalar and pseudo scalar mesons are studied in an IR-improved soft-wall AdS/QCD model. The Reissner-Nordstrom AdS black hole metric is introduced to describe both the finite temperature and density effects. The thermal spectral function is computed, which shows that the position of peak moves towards a smaller value and the width of peak also broadens to a larger value as the temperature increases. The critical temperature at which the peaks are completely dissolved has been found to be around $T_{c}\simeq 160\sim 200\MeV$. The effect of chemical potential is shown to be the same as the one caused by the temperature. It is found that when the temperature approaches to zero $T\approx 0$, the melting down of mesons occurs at the scaled critical chemical potential $\mu_0/\kappa\approx 2.2\sim 2.4 \GeV$.
\end{abstract}
\pacs{}

\maketitle

\section{Introduction}
\label{Chap:Intro}

The properties of thermodynamics of quantum chromodynamics (QCD), which relate to the experiments at the Relativistic Heavy Ion Collider (RHIC) and the Large Hadron Collider (LHC), have attracted a great deal of attention recently. The strongly correlated quark-gluon plasma (QGP) produced from RHIC involves many non-perturbative problems of QCD, such as chiral symmetry breaking and restoration, deconfinement phase transition, QCD phase diagram and so on. Many theoretical tools have been developed to describe those phenomena: effective QCD theories such as effective chiral dynamical model (CDM) \cite{Nambu:1960xd,DW,Huang:2011mx} of low energy QCD, QCD Sum Rules and lattice QCD. Although the effective CDM can provide well prediction for the mass spectra of ground state mesons, it is hard to characterize the linear trajectory behavior for the higher excited resonance meson states. When it comes to the dense matter situation, the lattice QCD calculation encounters difficulties for the sign problem in nonzero chemical potential case.

In the framework of the gauge/gravity duality approach \cite{Maldacena:1997re,Witten:1998zw,Gubser:1998bc}, the holographic QCD has been developed to bring new light on solving the problem of nonperturbative QCD. Searching the bulk gravity theory which corresponds to realistic QCD was first pursued in \cite{Witten:1998zw}. Mainly there are two complementary ways to follow: the top-down approach and the bottom-up approach. The former starts with the string solutions and find the brane configuration for the dual gravity to reproduce some basic QCD features, e.g. $D3/D7$ \cite{D3D7_1,D3D7_2,D3D7_3,D3D7_4}, $D4/D8/\overline{D8}$ \cite{D4D8_1,D4D8_2}, $D4/D6/\overline{D6}$ \cite{D4D6}. The latter known as a phenomenological AdS/QCD model consists of a gauge theory in a curved space with some field contents which are dual to some operators in QCD. The so-called hard-wall AdS/QCD model with a sharp infrared (IR) cut off $AdS_{5}$ metric was developed firstly in Refs. \cite{hardwall,hardwall1,hardwall2,hardwall3}. It can correctly realize the chiral symmetry breaking and low-lying mesons states, while the higher excited states of hadrons were found to deviate the Regge trajectory. Then the soft-wall AdS/QCD model \cite{soft-wall} was constructed to remedy this drawback by introducing a background dilaton field which suppresses the action gradually at IR region. Applying the WKB approximation analysis, it was shown that the dilaton's quadratic behavior at IR region will lead to the Regge trajectory for the excited meson spectra. However, the chiral condensation in the soft-wall model was found to be proportional to quark mass. A large amount of works have been done to incorporate both the chiral symmetry breaking and the linear trajectory of excited meson spectra in the AdS/QCD model. In \cite{quartic term}, a quartic interaction term has been introduced into the action in soft-wall model to realize the explicit and spontaneous chiral symmetry breaking, while such a term was shown to cause an instability of the scalar potential and a negative mass for the lowest lying scalar meson state. In Refs. \cite{pADSQCD1,pADSQCD2}, by simply modifying the bulk gravity at IR region, a consistent soft-wall AdS/QCD model has been constructed and the resulting mass spectra for light resonance states have been found to agree with the experimental data.

It is also interesting to investigate the finite temperature and density properties in  the AdS/QCD model. The hard-wall's finite temperature effects was studied in \cite{hardwall_T}. In Ref. \cite{Scalar glueball,glueball_1,glueball_2}, the scalar glueball and light scalar mesons spectral have been studied in the soft-wall model and the highest temperature for which the meson state can exist was found to be about $40-60\MeV$. Compared with the Hawking-Page (HP) transition temperature in the soft-wall model, which was found to be $T_{HP}\backsimeq192\MeV$ in \cite{HPtransition}, the meson states dissociation occurs in the confined QCD phase, which is far from the deconfinement transition. One way to increase the temperature is to set up a different mass scale in the soft-wall model. In Ref. \cite{jpsi1,jpsi2}, $J/\psi$ thermal spectral function was studied by treating the scale as a flavor-dependent parameter. In Ref. \cite{jpsi3}, the melting of $J/\psi$ spectral peak was found to occur at $T\approx540\MeV$. However, the light mesons' dissociation temperature is still too low. The chemical potential is another interesting properties, which has been widely investigated in holographic QCD. In Ref. \cite{D3D7Chemical_1,D3D7Chemical_2,D3D7Chemical_3}, the spectral functions have been studied by taking D3/D7 setup with fundamental matters at finite baryon density. Density effects on the spectral function in the soft-wall model have been discussed in \cite{Chemical_spectral_1,Chemical_spectral_2}.

In \cite{VandAV}, we extended the the IR-improved soft-wall \cite{pADSQCD1,pADSQCD2} to a thermodynamic model at finite temperature. Spectral functions of vector and axial-vector mesons have been computed. The highest temperature for which the meson state can survive was found to be around $T_{c}\simeq200\MeV$ without quartic interaction term. However, difficulty remains in obtaining the spectral function of pseudo-scalar meson due to the coupling between the axial-vector field and the pseudo-scalar field. In this paper, we only pay attention to the pseudoscalar and scalar mesons. We construct the soft-wall AdS/QCD model which works well in both zero temperature and finite temperature region for the pseudoscalar and scalar mesons. In Sec. \ref{Chap:Model}, inspired by Refs. \cite{pADSQCD1,pADSQCD2}, we will build a thermodynamic model for scalar part by introducing a modified Reissner-Nordstrom AdS black hole. In Sec. \ref{Chap:zero temperature}, parameters  are fixed at zero temperature with density. The effects of finite temperature and density will be discussed in Sec. \ref{Chap:Finite T}. Our conclusions and remarks are presented in the final section.

\section{Improved soft-wall AdS/QCD Model}
\label{Chap:Model}

In this section, we will construct a phenomenological thermodynamic AdS/QCD model inspired by the predictive holographic QCD model \cite{pADSQCD1,pADSQCD2}. In Ref.\cite{VandAV}, we studied the spectral functions of the vector and axial-vector mesons. However, the pseudo-scalar part is the most intriguing one due to the coupling with axial-vector field. As our main goal is to investigate the finite temperature effects, we will consider the  simple case by taking off the gauge field part and focus only on scalar and pseudo-scalar part in the 5D action. Thus the 5D action becomes:
\begin{equation}
S=\int d^{5}x\,\sqrt{g}e^{-\Phi(z)}\,{\rm {Tr}}\left[|D_{M}X|^{2}-m_{X}^{2}|X|^{2}-\lambda|X|^{4}\right],\label{action}
\end{equation}
with $m_{X}^{2}=-3$. The quartic term $\lambda|X|^{4}$ is introduced to improve the ground state of scalar mesons masses which have been discussed in  Ref.\cite{pADSQCD1}. $\Phi$ is the dilaton background field, which can lead to a linear trajectory if the UV behavior is taken as $\Phi(z\rightarrow\infty)\sim z^{2}$\cite{soft-wall}. The fundamental field $X$, which corresponds to a dimension three scalar operator $\bar{q}_{R}^{\alpha}q_{L}^{\beta}$, can be written as: $X(x,z)\equiv\left( X(z) +S(x,z)e^{2i\pi(x,z)}\right)$. $S(x,z)$ and $\pi(x,z)$ are scalar field and pseudo-scalar field respectively, $X(z) = \left\langle X(x,z)\right\rangle $ is the bulk vacuum expectation value.

In order to investigate finite temperature and density effects, we need to find a dual geometry in the bulk. According to AdS/CFT correspondence, the quark chemical potential, which is the coefficient of the quark number operator $\bar{q}\gamma^{0}q$, corresponds to the time-component of $U(1)$ gauge field in the bulk. The gravity action which describes the interaction of the $U(1)$ gauge field with the AdS space is given by:
\begin{equation}
S=\int d^{5}x\sqrt{-g}\left[\frac{1}{2\kappa^{2}}(\mathcal{R}-2\Lambda)-\frac{1}{4g^{2}}F^{MN}F_{MN}\right],\label{gravity action}
\end{equation}
When using the following ansatz:
\begin{eqnarray}
& & A_{t}=A_{t}(z),\\
& & A_{i}=0 \qquad\qquad (i=1,\cdots,3,z),\\
& & ds^{2}=\frac{R^{2}}{z^{2}}\left(f(z)dt^{2}-d\vec{x}^{2}-\frac{dz^{2}}{f(z)}\right),\label{ansatz}
\end{eqnarray}
which leads to a well-known solution called Reissner-Nordstrom AdS black hole:
\begin{eqnarray}
& & f\left(z\right)=1-\left(1+Q^{2}\right)\left(\frac{z}{z_{h}}\right)^{4}+Q^{2}\left(\frac{z}{z_{h}}\right)^{6},\\
& & A_{t}\left(z\right)=\mu-\kappa\frac{Q}{z_{h}^{3}}z^{2}, \label{RN solution}
\end{eqnarray}
with $0\leqslant Q\leqslant\sqrt{2}$, which is the charge of the gauge field.

In this paper, we will adopt a modified Reissner-Nordstrom AdS black hole inspired by the one introduced in \cite{VandAV}.
\begin{equation}\label{RN deformed metric}
ds^2 = a^2(z) \left( f(z) dt^2 - \sum_{i=1}^3 dx^2_i - \frac{dz^2}{f(z)} \right),
\end{equation}
In order to incorporate both chiral symmetry breaking and linear confinement, the IR region of the metric is modified as:
\begin{eqnarray}\label{a(z)}
a^2(z) = 1/z^2 + \mu_g^2
\end{eqnarray}
Here $\mu_g$ is a constant mass scale which characterizes QCD confinement. As shown in  \cite{pADSQCD1,pADSQCD2}, such an improved 5D metric will lead to a non-trivial dilation solution.  $f(z)$ is the same as the one in Eq.~\ref{RN solution}, so the Hawking Temperature, which corresponds to the temperature in boundary theory, is not changed under this modification:
\begin{eqnarray}\label{hawking T}
T_{H}=\frac{1}{4\pi}\left|\frac{df}{dz}\right|_{z\rightarrow z_{h}}=\frac{1}{\pi z_{h}}\left(1-\frac{Q^{2}}{2}\right)
\end{eqnarray}
The chemical potential can be defined by the condition that gauge field vanishes near the black hole horizon: $A_{t}\left(z_{h}\right)=0$, so we have
\begin{eqnarray}\label{def of mu}
\mu=\kappa\frac{Q}{z_{h}}
\end{eqnarray}
$\kappa$ is a free parameter, which can be determined in many different ways \cite{Dual Geometry of the Hadron,HQCD phase}. In this paper, we will directly give the result of $\mu/\kappa$.

 The bulk vacuum expectation value of the fundamental field $X$ has the following form for the two flavor case
\begin{eqnarray}\label{X(z)}
\left\langle X(z) \right\rangle = \frac{1}{2}\,v(z)~\mathbf{1}_2,
\end{eqnarray}
While the VEV $v(z)$ is related to the dilaton field $\Phi(z)$ through the equation of motion:
\begin{eqnarray}
0=\partial_{z}\Big(a^{3}(z)f(z)e^{-\Phi(z)}\partial_{z}v(z)\Big)-a^{5}(z)e^{-\Phi(z)}\left(m_{X}^{2}+\frac{\lambda}{2}\right)v(z).\label{equation of VEV}
\end{eqnarray}

In this paper, we take the same form of VEV $v(z)$ proposed in \cite{VandAV}. It consists of two parts: zero temperature part $v_0(z)$ and the finite temperature part $v_1(z) \ln f(z)$.
\begin{eqnarray}\label{vassume}
v(z) = v_1(z) \ln f(z) + v_0(z),
\end{eqnarray}
where the zero temperature part $v_0(z)$ is taken from the type IIb model in Ref.\cite{pADSQCD1} which is given by Table.~\ref{TBofv0}. Its IR boundary behavior $v_{0}(z\rightarrow\infty)\sim\sqrt{z}$ can lead to a more accurate prediction for all the mass spectra in zero temperature and density.

\begin{table}[!h]
\begin{center} \begin{tabular}{ c l l l l }
\hline\hline
Model & $\qquad\qquad$ $v_0(z)$  & \multicolumn{3}{c}{Parameters}   \\
\hline
IIb & $ z(A+B z^2)(1+C z^4)^{-5/8}$ & $A = m_q \zeta$,& \quad $B=\sigma/{\zeta}$,       & \quad  $C=(B^2/(\mu_d\gamma^2))^{4/5}$ \\
\hline\hline
\end{tabular}
\caption{IIb model which is taken from \cite{pADSQCD1} with $\mu_g=\mu_d/\sqrt{3}$. The numerical values are shown in Tabel.~\ref{Table:parameter}. }
\label{TBofv0}
\end{center}
\end{table}

The finite temperature part is introduced to regulate the singular behavior of $\Phi(z)$ near the horizon \cite{VandAV}. From Eq.~\ref{equation of VEV} , we can obtain the non-trival form of the dilaton. In the vicinity of horizon, dilaton field $\Phi(z)$ has a divergent behavior:
\begin{equation}
\Phi^{'}(z)=\frac{\ln f(z)}{v_{1}(z)f^{'}(z)}\mathcal{F}(z)+\cdots\label{dilaton}
\end{equation}
with
\begin{equation}
\mathcal{F}(z)\equiv3a^{2}(z)v_{1}(z)+f^{'}(z)v_{1}^{'}(z)\label{F}
\end{equation}
Note that near the horizon, $f(z_{h})\rightarrow0$, the term $\ln f(z)$ becomes divergent. If $v_{1}(z)$ takes the form as the solution of $\mathcal{F}(z)=0$, the dilaton gets smooth solution at the horizon as shown in \cite{VandAV}. In this work, $v_{1}(z)$ becomes too complicated due to the introduction of chemical potential. In zero density limit, $v_{1}(z)$ has the same form as in \cite{VandAV}.
\begin{eqnarray} \label{v1}
v_1(z) = c_v \exp\left[  \frac{m_X^2}{8(\pi T)^4z^4} \left( \frac{1}{2} + \mu_g^2 z^2 \right)\right],
\end{eqnarray}
Notice that $v_{1}$ vanishes when temperature approaches zero due to the minus sign of $m_{X}^2$. When considering the effects of the quartic interaction term, the coefficient of the quartic term $\lambda$ takes the following form \cite{VandAV}
\begin{eqnarray}\label{lambdaT}
\lambda = \frac{\lambda_0}{1+ c_\lambda (\ln f(z))^p} \qquad {\rm with} \qquad p \ge 3,
\end{eqnarray}
For simplicity, we still take $p=3$ in current consideration. Thus all parameters in this model will be inputted  by experimental data at zero temperature and density limit in next section.

\section{Model at Zero Temperature and Density Limit}
\label{Chap:zero temperature}

\subsection{Fitting Parameters}
\label{parameters input}

We will make numerical calculations for the mass spectra of scalar and pseudo-scalar mesons in zero temperature and density limit. Turning off chemical potential and temperature, which means $f(z)\to1$ and $Q\to0$, the metric takes the following form which is the same as the one used in \cite{pADSQCD1,pADSQCD2}.
\begin{equation}\label{metric_dads_01}
ds^2 = a^2(z) \left(  dt^2 - \sum_{i=1}^3 dx^2_i - dz^2 \right),
\end{equation}
The finite temperature part of VEV vanishes, leaving only the zero temperature part.
\begin{eqnarray}\label{vassume}
v(z) = v_0(z),
\end{eqnarray}
$\mu_g$ as a mass scale can be fixed from a global fitting. The left three parameters $m_q$, $\sigma$ and $\gamma$ are fixed by the well measured experimental value of $m_\pi=139.6\MeV$, the Gell-Mann-Oakes-Renner relation $f_{\pi}^2m_{\pi}^2 = 2m_q \sigma$, and the mass of scalar mesons. The values of three fitting parameters are presented in Table.~\ref{Table:parameter}.

\begin{table}[!h]
\begin{center}
\begin{tabular}{ccccc}
\hline\hline
$\lambda_0$ & $m_q$ (MeV) & $\sigma^{\frac{1}{3}}$ (MeV) & $\mu_d$ (MeV)  & $\gamma$ \\
\hline
$\lambda_0=0$ & 4.30 & 268 & 445    & 0.1\\
$\lambda_0=15$ & 6.99 & 228 & 505    & 0.124\\
\hline\hline
\end{tabular}
\caption{The numerical values of parameters for the case without the quartic term $(\lambda_0=0)$ and the ones with the quartic term $(\lambda_0=15)$, with $\zeta=\sqrt{3}/(2\pi)$ and $\mu_g=\mu_d/\sqrt{3}$.}
\label{Table:parameter}
\end{center}
\end{table}

\subsection{Mass Spectra of Scalar and Pseudo-Scalar Mesons}

The fundamental scalar field can be decomposed as $X(x,z)\equiv(v(z)/2+S(x,z))e^{2i\pi(x,z)}$, where $S(x,z)$ is the
scalar meson field and $\pi(x,z)=\pi^a(x,z) t^a$ the pseudo-scalar field. Now we can derive the equations of motion for both two fields:
\begin{eqnarray}\label{asu2}
\textrm{S}&:& S_{n}^{''}(z)+S_{n}^{'}(z)\left(\frac{3a^{'}(z)}{a(z)}-\Phi^{'}(z)\right)+S_{n}(z)\left(m_{S_{n}}^{2}+3a(z)^{2}+\frac{3}{2}\lambda a(z)^{2}v(z)^{2}\right)=0\\
\textrm{PS} &:& \pi_{n}^{''}(z)+\pi_{n}^{'}(z)\left(\frac{3a^{'}(z)}{a(z)}+\frac{2v^{'}(z)}{v(z)}-\Phi^{'}(z)\right)+m_{\pi}^{2}\pi_{n}(z)=0,\\
\end{eqnarray}

Under the boundary condition $S_n(z\rightarrow0)=0, \p_z S_n(z\rightarrow\infty)=0;\pi_n(z\rightarrow0)=0, \p_z \pi_n(z\rightarrow\infty)=0 $, we numerically solve the equations by shooting method. In this way, we find out the normalizable modes for scalar and pseudo-scalar fields in the bulk and their eigenvalues, which are the masses of mesons. The predictive mass spectra for the scalar and pseudo-scalar mesons are showed in Table.~\ref{mass}.  Note that the scalar states $f_0(980\pm10)$, $f_0(1505\pm6)$, $f_0(2103\pm8)$, $f_0(2314\pm25)$ should be classified into the isosinglet resonance scalar states of $SU(3)$ octet mesons, rather than the $SU(3)$ singlet resonance scalar states. Here to circumvent the difficulty of the finite behavior of pseudoscalar mesons, we have ignored the coupling between pseudoscalar field and axial-vector field from the action \ref{action}. This will cause the tiny mass splittings between scalar and pseudo-scalar meson mass eigenvalues, especially the ground state mesons. After adding the quartic interaction term, mass spectra for scalar meson can be improved a little.

\begin{table}[!h]
\begin{center}
\begin{tabular}{c c c c | ccc }
\hline\hline
  n & $\pi$~experimental.~(MeV) &   $\lambda=0$ &   $\lambda=15$ &  $f_0$~experimental.~(MeV) & $\lambda=0$ & $\lambda=15$  \\
  \hline 
  0 &      139.6            &  139.6    &     139.6      &                $550^{+250}_{-150}$  &   139.8         &     275.3           \\
  1 &   $1350 \pm 100$      &  1439    &     1431      &                      $1350 \pm 150$  &   1439     &    1453               \\
  2 &   $1816 \pm 14$       &   1701     &    1738     &                   $1724 \pm 7$      &    1701     &    1758             \\
  3 &                        &  1926     &   1967       &                   $1992 \pm 16$     &   1926      &   1987             \\
  4 &                        &    2126    &   2159     &                 $2189 \pm 13$       &   2126     &     2179             \\
  5 &                         &   2308   &   2326      &                                     &   2476      &    2346            \\
  5 &                         &  2475    &   2476      &                                     &    2633     &    2495             \\
\hline \hline
\end{tabular}
\end{center}
\caption{The experimental and the predicted mass spectra for Scalar (right side) and Pseudo-Scalar (left side) meson}
\label{mass}
\end{table}

\section{Model with Finite Temperature and Density}
\label{Chap:Finite T}

When temperature and chemical potential are turned on, the equation of motion is given as follows in momentum space by performing the Fourier transformation(we put three-momentum to zero $\overrightarrow{p}=0$):
\begin{eqnarray}
\label{eq:eomST}
\textrm{S}&:&
K''(z)+K'(z) \left(\frac{3
   a'(z)}{a(z)}+\frac{f'(z)}{f(z)}-\Phi '(z)\right)\nonumber
   \\ && \qquad\qquad\qquad\quad +K(z)\left(\frac{3 \lambda  a(z)^2 v(z)^2}{2 f(z)}+\frac{3 a(z)^2
   f(z)+\omega ^2}{f(z)^2}\right)=0,\\
\label{eq:eomPST}
\textrm{PS}&:&
K''(z)+K'(z) \left(\frac{3 a'(z)}{a(z)}+\frac{f'(z)}{f(z)}+\frac{2
   v'(z)}{v(z)}-\Phi '(z)\right)
 +\frac{\omega ^2 K(z)}{f(z)^2}=0,
\end{eqnarray}

As the temperature and density increase, the horizon of black hole $z_{h}$ moves from infinity to boundary side. Solutions of equations of motion drop into black hole before they vanish. Therefore we cannot use the method of finding eigenmode. Alternatively,  we will investigate spectral function which is the imaginary part of the retarded Green's function.

First we will check the boundary behavior of the solution. Near the origin, we can extract the asymptotic solution of Eq.~\ref{eq:eomST} and Eq.~\ref{eq:eomPST}. For convenience, we change the radial coordinate $z$ to dimensionless variable $u$ as $u=z/z_h$. The two linear independent solutions are given as follows:
\begin{eqnarray}\label{asym solu}
\textrm{S} &:&
K _1\to u^2
   J_1\left(\frac{u \sqrt{\frac{3
   \lambda  A^2}{2}+\omega ^2+3 \mu
   _g^2}}{z_{h}}\right),
\quad
K _2\to u^2
   Y_1\left(\frac{u \sqrt{\frac{3
   \lambda  A^2}{2}+\omega ^2+3 \mu
   _g^2}}{z_{h}}\right)  \nonumber
 \\
\textrm{PS} &:&
K _1\to u
   J_1\left(\frac{u \omega }{z_{h}}\right),
\quad
K _2\to u
   Y_1\left(\frac{u \omega }{z_{h}}\right)
\end{eqnarray}
Here, $J_1$ and $Y_1$ are the first-kind Bessel function and second-kind Bessel function respectively. On the other hand, near the horizon, we take the in-falling boundary condition which corresponds to retarded Green's function \cite{retarded green}:
\begin{eqnarray}\label{infalling}
K_{-}\to(1-u)^{-i\frac{z_{h}\omega}{2\left(2-Q^{2}\right)}}
\end{eqnarray}
The solutions of equation of motions can be expressed by the combination of the two independent asymptotic solutions. The coefficients $A(\omega, q)$ and $B(\omega,q)$ will be fixed by the IR in-falling boundary condition:
\begin{eqnarray}
K(u)=A(\omega,q)K_{1}(\omega,q,u)+B(\omega,q)K_{2}(\omega,q,u)\longrightarrow(1-u)^{-i\frac{z_{h}\omega}{2\left(2-Q^{2}\right)}}\label{solution}
\end{eqnarray}

The on shell action of scalar field part reduces to surface terms:
\begin{equation}
S=\int\frac{d^{4}p}{\left(2\pi\right)^{4}}\left.e^{-\Phi(z)}f(z)a(z)^{\frac{3}{2}}S(p,z)\partial_{z}S(p,z)\right|_{z=0}^{z=z_{h}},\label{surface}
\end{equation}
Following the prescription in \cite{retarded green}, after substitute Eq.~\ref{asym solu} into surface terms, we find that the spectral function which relates to the imaginary part of two point retarded Green function is proportional to the imaginary part of $B(\omega,q)/A(\omega, q)$. The results for pseudo-scalar part is the same.
\begin{eqnarray}
\rho(\omega,q)=-\frac{1}{\pi}\mathrm{Im}\, G(\omega,q)\,\theta(\omega^{2}-q^{2})\varpropto\mathrm{Im}\,\frac{B(\omega,q)}{A(\omega,q)},
\end{eqnarray}

Let us now make a numerical calculation. As we have discussed in \cite{VandAV}, $c_v$ and $c_\lambda$ play the role as regulators to regulate the singular behavior near the horizon, which have no effect on finite temperature and density behavior. In principle, we should assign their values as small as possible, while it turns out to be more difficult if we adopt extremely small values in practical calculation. In Ref. \cite{Cui:2011wb}, we have computed the quark number susceptibility at finite temperature by taking the limit that momentum and frequency are eventually taken to be zero. In this case, the modified terms which contain $c_v$ and $c_\lambda$ can be neglected. In this paper, we take $c_v=c_\lambda=10^{-3}$.

\subsection{Critical Behavior with Finite Temperature ( $T\neq0$ and $\mu=0$)}

\begin{figure}[!h]
\begin{center}
\includegraphics[width=64mm,clip]{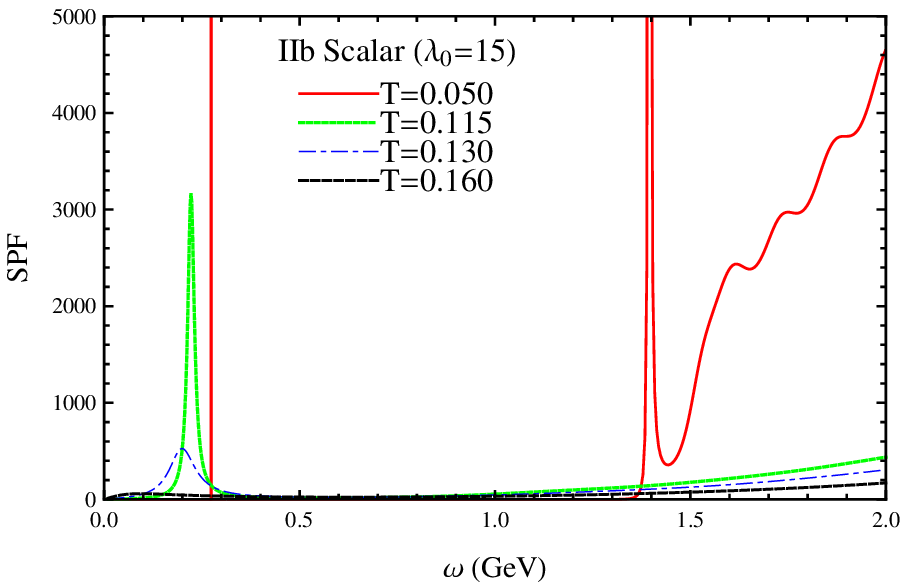}
\includegraphics[width=64mm,clip]{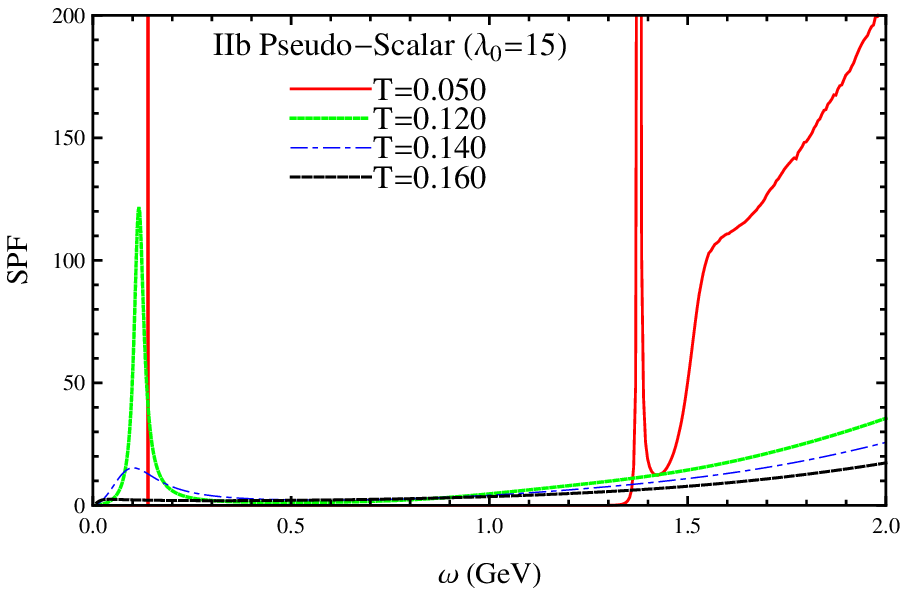}
\includegraphics[width=64mm,clip]{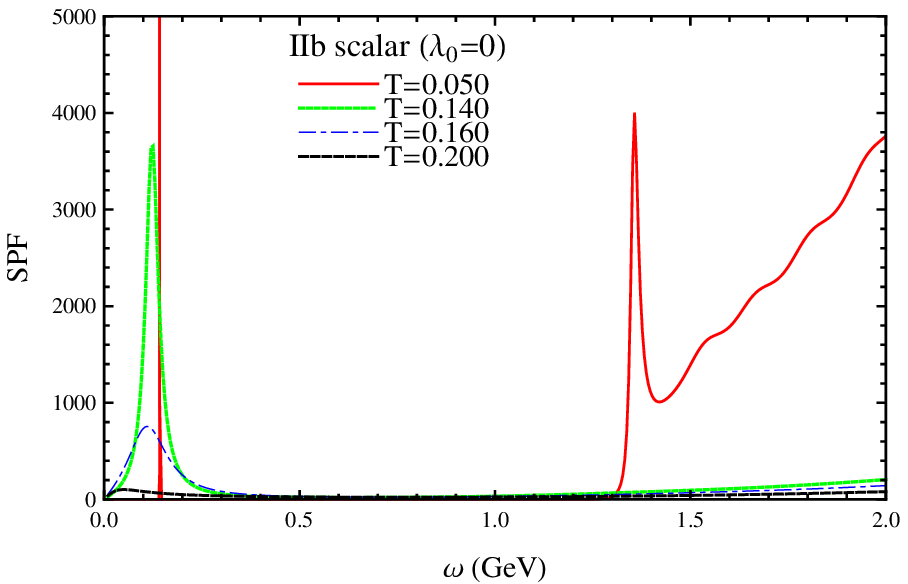}
\includegraphics[width=64mm,clip]{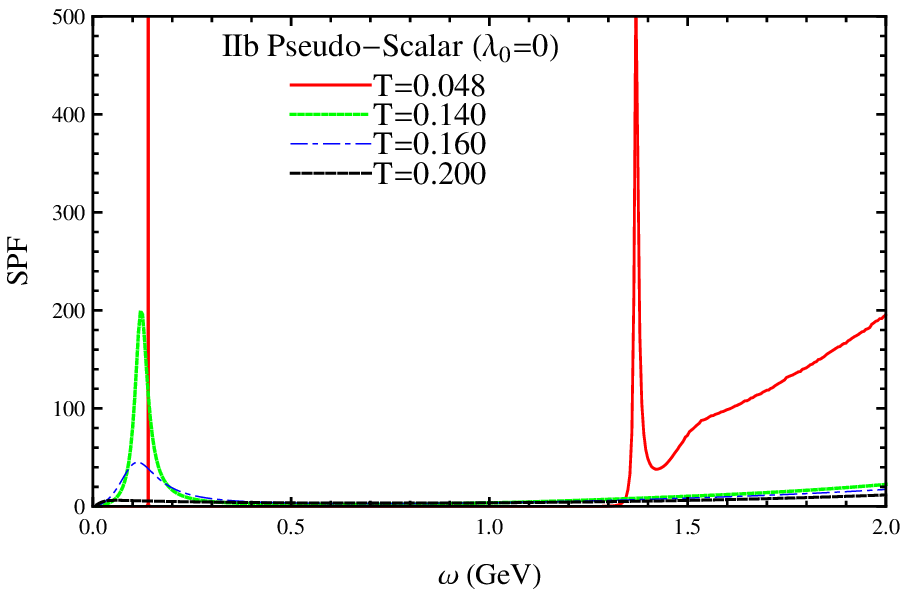}

\end{center}
\caption{The results of spectral function for scalar meson (left side) and pseudo-scalar meson (right side).
}\label{Fig lamda}
\end{figure}

In Fig.~\ref{Fig lamda}, we turn off chemical potential and check the finite temperature behavior in scalar and pseudo-scalar channel. As the temperature increases, the horizon $z_{h}$ approaches boundary so that both scalar and pseudo-scalar mesons become unstable and melt. For the low temperature, the positions of the peaks are in accord with the masses which have been given in zero temperature case in Table.~\ref{mass}. Increasing temperature, the peaks are shifted towards smaller values and the widths become broader. We can define a melting point, where no peak can be distinguished above this temperature. Quantitatively, melting temperature can be determined from the Breit-Wigner form as showed below in Eq.~\ref{BWform}.That is if the continuum part $P(\omega^2)$ is larger than the height of the peak, then it can be considered that no mesons state exist. So the temperature $T_c\simeq 200\MeV(\lambda=0)$ or $T_c \simeq 160\MeV$($\lambda=15$) can be taken as the critical temperature of dissolution of mesons state. Here the critical temperature $T_c \simeq 200\MeV$ without the quartic term is consistent with other NJL models' predictions \cite{Huang:2011mx,Tmuplane,NJL1,NJL2,NJL3}, and also with the one obtained from the melting point for thermal mass spectra of the vector and axial-vector mesons\cite{VandAV} . The value $T_c \simeq 160\MeV$ with the quartic term approaches to the critical temperature yielded from the quark number susceptibility when the phase changes from the hadron phase to the quark-gluon-plasma phase at zero chemical potential\cite{Cui:2011wb}, and also to the one from the lattice QCD calculations\cite{LQCD}.

\begin{figure}[!h]
\begin{center}
\includegraphics[width=64mm,clip]{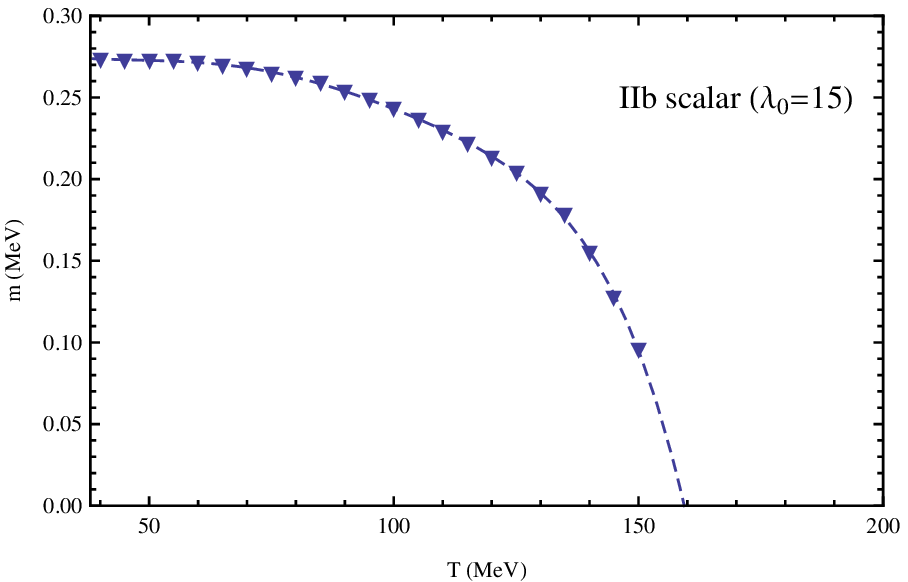}
\includegraphics[width=64mm,clip]{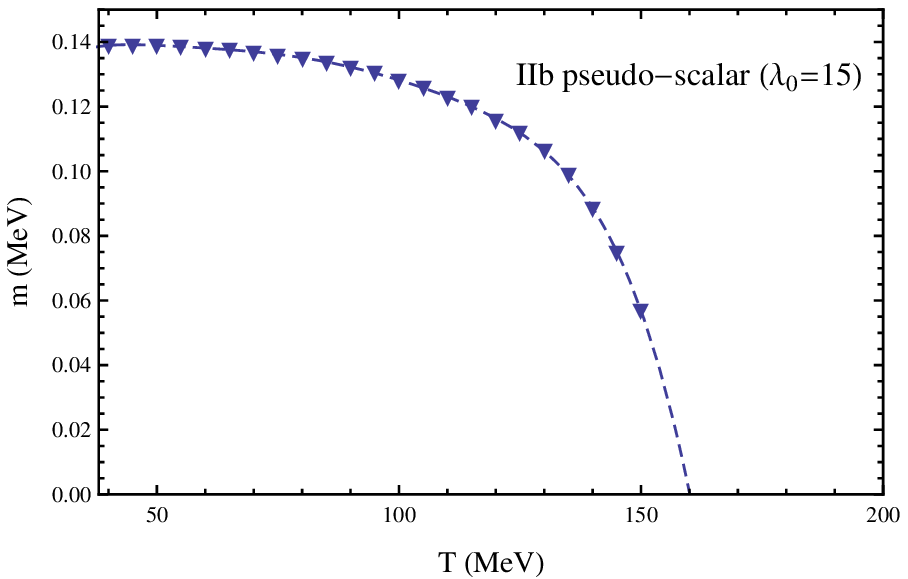}

\end{center}
\caption{The relation between the location of the first peak and the temperature for both scalar and pseudo-scalar parts with quartic interaction term.
}\label{mass shift}
\end{figure}

We can get more information by fitting the spectral function with a Breit-Wigner form:
\begin{eqnarray}\label{BWform}
\frac{a \omega^b}{(\omega^2-m^2)^2+\Gamma^2} + P(\omega^2).
\end{eqnarray}
where $m$ and $\Gamma$ correspond to the location and width of the peak. $P(\omega^2)$ is representing a continuum which is taken the form $P(\omega^2)=c_1 +c_2 \omega^2 +c_3 (\omega^2)^{c_4}$. To see the shift of the mass of resonance in both scalar and pseudo-scalar channels, we consider the lowest lying state with quartic term and plot the relation between the mass $m$ and the temperature $T$ in Fig.~\ref{mass shift}.

It can be seen from Fig.~\ref{mass shift} that in the low temperature region the location of the peak drops linearly as temperature increases. When temperature is near $T=100\sim150 \MeV$, the mass starts to decrease drastically, and at $T=150\MeV$ the mass is reduced to about $30\%$ of its value at $T=0$. The lower the temperature is, the more difficult the numerical analysis is. This is because the horizon $z_{h}$ becomes too large and the method cannot be used anymore at very low temperature. We have to adopt the way used in zero temperature calculation to find the eigenvalue of the solution. In this work, we shall not investigate such a region intensively. Just from the extrapolation of the fitted curve, we can find that around the critical temperature $T_c \simeq 160\MeV$ the mass seems to approach zero. The same results can also be found in some other model's prediction \cite{Huang:2011mx}, which has been shown to relate to the chiral symmetry restoration at finite temperature.

\subsection{Effect from Chemical Potential ($T\neq0$ and $\mu\neq0$)}

\begin{figure}[!h]
\begin{center}
\includegraphics[width=64mm,clip]{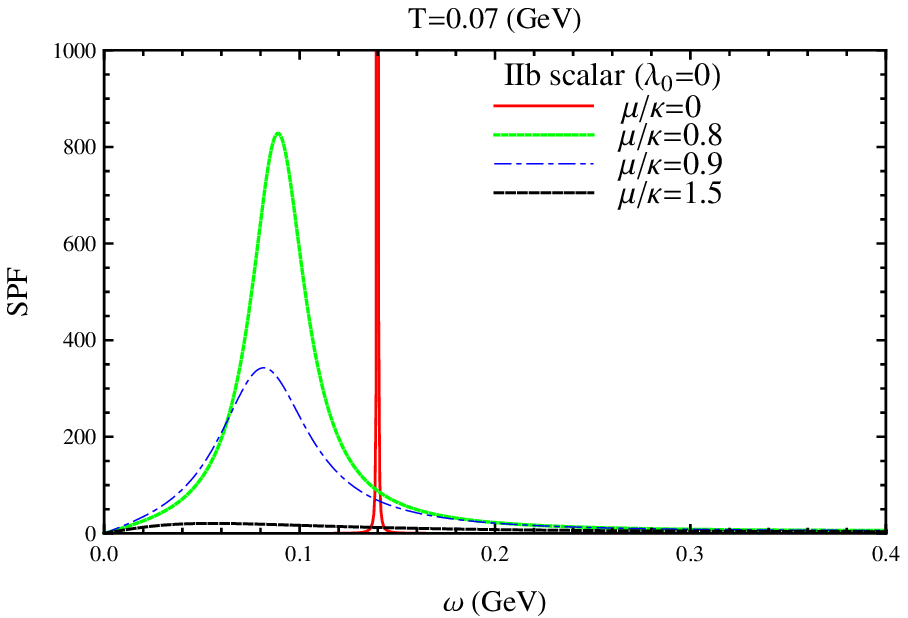}
\includegraphics[width=64mm,clip]{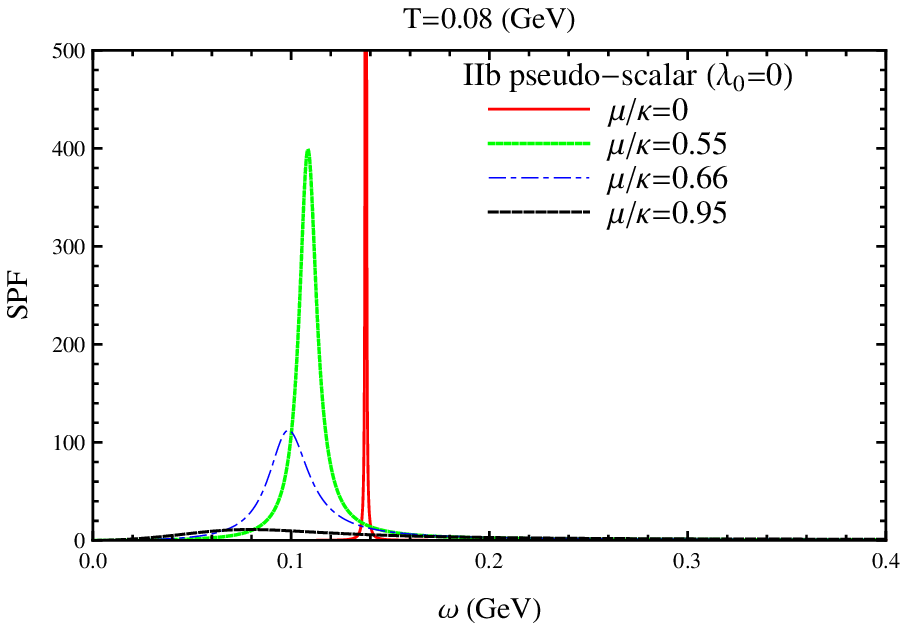}

\end{center}
\caption{The chemical potential effects on spectral functions for both scalar and pseudo-scalar mesons.
}\label{chemical potential}
\end{figure}

Let us now turn on chemical potential. From Eq.~\ref{def of mu}, it can be seen that the effect of chemical potential is the same as the temperature and makes horizon $z_{h}$ move close to the boundary side. In Fig.~\ref{chemical potential}, we present the numerical results for the first peak of the scalar and pseudo-scalar mesons spectral function in different chemical potential. Here we have used the ratio $\mu/\kappa$ instead of fitting the dimensionless parameter $\kappa$. It is shown that the increasing of chemical potential leads to the same behavior as the temperature does. As the horizon moves towards boundary, the solutions are absorbed by the black hole and the bound state starts to collapse.

\begin{figure}[!h]
\begin{center}
\includegraphics[width=64mm,clip]{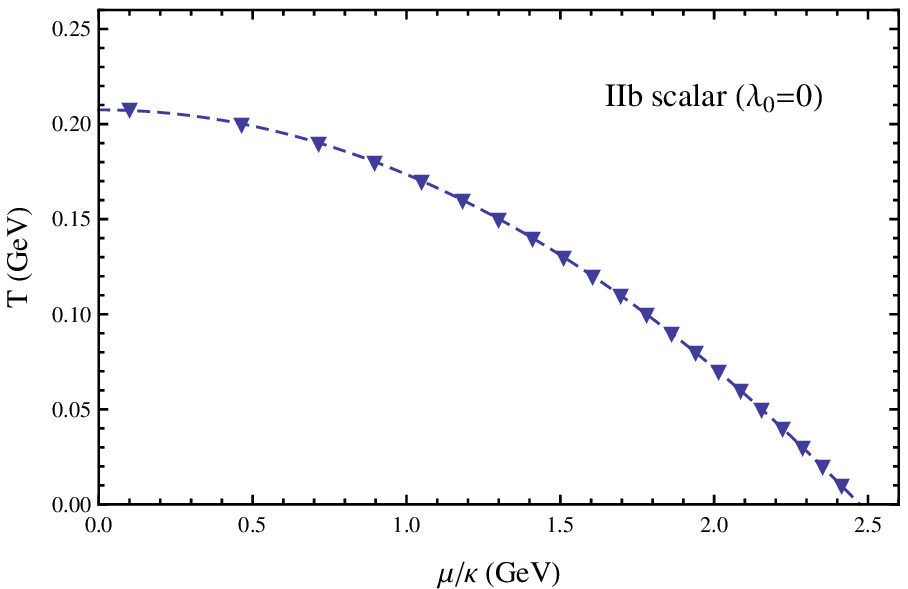}
\includegraphics[width=64mm,clip]{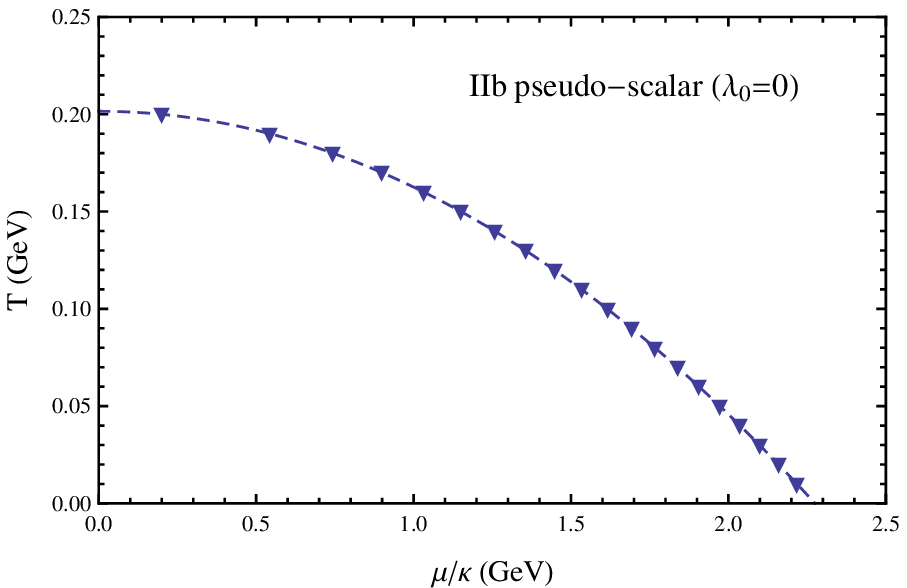}

\end{center}
\caption{Critical points of scalar and pseudo-scalar parts in $T-\mu$ planes}\label{Tmuplane}
\end{figure}

As the increasing of both temperature and density can lead to an unstable state as expected, we can define a critical point where the first peak of ground state completely melt down and cannot be distinguished anymore. Finding out the critical values $(T_c,\mu_c)$ for both temperature and density, we can plot it in the $T-\mu$ plane. The critical points for both scalar and pseudo-scalar parts are showed in Fig.~\ref{Tmuplane}. When the temperature approaches to zero ($T\to 0$), the critical chemical potential is found to be $\mu_{0}/\kappa=2.4\GeV$ for scalar part and $\mu_{0}/\kappa=2.2\GeV$ for pseudo-scalar part. On the other hand, when the chemical potential goes to zero $\mu=0$, one obtains the critical temperature $T_{0}\approx200\MeV$ which is the same as the one given in previous subsection. The curve in the $T-\mu$ plane divides the whole plane into two regions. The region under the curve represents the phase in which the meson states still exist, which can be interpreted as the "hadronic phase". While the region above the curve corresponds to the real QCD's quark-gluon phase in which  the meson states completely dissolve. As mesons are the bound states of quarks and antiquarks,  the critical behavior of their disappearance can be understood as the chiral symmetry restoration.

For the free dimensionless parameter $\kappa$, one may get its value by comparing with other models' results. In \cite{Tmuplane}, there appears a relation $T_{0}/\mu_{0}\approx 1/2 $ for the critical chemical potential $\mu_0$ at $T=0$ and  critical temperature $T_0$ at $\mu=0$, which then allows one to fix the parameter $\kappa\approx 1/6$.

\section{Conclusions and remarks}
\label{Chap:Sum}

We have investigated the IR-improved soft-wall AdS/QCD model with zero temperature and finite temperature. An IR-improved Reissner-Nordstrom AdS black hole metric has been introduced to describe the finite temperature and density effects. Following \cite{VandAV}, an additional finite temperature part of bulk vacuum expectation value is introduced to get a smooth dilaton solution. In zero temperature case, the predictive resonance meson states agree well with the experimental data except for the ground state of scalar mesons which will further be investigated elsewhere.  At finite temperature, we have computed the spectral function of mesons, a broadening of the peaks and moving towards the smaller values of the mass as the temperature increases have been demonstrated in detail. The critical temperature at which the peaks completely dissolved has been found to be around $T_{c}\simeq 200\MeV$(without quartic term) and $T_{c} \simeq 160\MeV$(with quartic term). By fitting the spectral function in terms of the Breit-Wigner form, we have quantitively studied the mass shift as the temperature increases. The feature that pion mass approaching to zero around the critical temperature agrees well with many other theoretical predictions. We have also investigated the effect of chemical potential. Its effect has been shown to be the same as the one caused by the temperature, namely it can lead to an unstable meson sate. It has been found that when the temperature goes to zero $T\approx0$, the melting down of mesons occurs at the scaled chemical potential $\mu_0/\kappa\approx 2.2 \GeV$ for pseudo-scalar mesons and $\mu_0/\kappa\approx 2.4\GeV$ for scalar mesons.

\section*{Acknowledgements}
This work is supported in part by the National Nature Science Foundation of China (NSFC) under Grants No. 10975170, No.10905084,No.10821504; and the Project of Knowledge Innovation
Program (PKIP) of the Chinese Academy of Science. \\

\appendix

\end{document}